\documentclass[a4paper,fleqn,usenatbib]{mnras}

\usepackage{newtxtext,newtxmath}

\usepackage[T1]{fontenc}
\usepackage{ae,aecompl}

\usepackage{graphicx}	
\usepackage{amsmath}	
\usepackage{amssymb}	

\usepackage[utf8]{inputenc}
\usepackage{natbib}
\usepackage{threeparttablex} 
\usepackage{booktabs} 
\usepackage{longtable}
\usepackage{rotating}
\usepackage{amsbsy}
\usepackage{url} 
\usepackage{ulem}

\title[The Colour Magnitude Diagram of Crater~II]{A DECam View of the Diffuse Dwarf Galaxy Crater~II: The Colour-Magnitude Diagram}

\author[A.R. Walker et al.]{
A.R. Walker,$^{1}$\thanks{E-mail: awalker@ctio.noao.edu}
C.E. Mart\'inez-V\'azquez,$^{1}$
M. Monelli,$^{2,3}$
A.K. Vivas,$^{1}$
G. Bono,$^{4}$
\newauthor
C. Gallart,$^{2,3}$
S. Cassisi,$^{5,6}$
G. Andreuzzi,$^{7,11}$
E.J. Bernard,$^{8}$
M. Dall'Ora,$^{9}$
\newauthor
G. Fiorentino,$^{10,11}$
D.L. Nidever,$^{12,13}$
K. Olsen,$^{12}$
A. Pietrinferni,$^{5}$
P.B. Stetson$^{14}$
\\
$^{1}$ Cerro Tololo Inter-American Observatory, NSF's National Optical-Infrared Astronomy Research Laboratory, Casilla 603, La Serena, Chile\\
$^{2}$ Instituto de Astrof{\'\i}sica de Canarias, Calle V{\'\i}a L\'actea, E-38205 La Laguna, Tenerife, Spain\\
$^{3}$ Universidad de La Laguna, Dpto. Astrof{\'\i}sica, E-38206 La Laguna, Tenerife, Spain\\
$^{4}$ Departimento di Fisica, Universit\'{a} di Roma Tor Vergata, via della Ricerca Scientifica 1, 00133, Rome, Italy\\
$^{5}$ INAF-Osservatorio Astronomico d'Abruzzo, Via M. Maggini, I-64100 Teramo, Italy\\
$^{6}$ INFN-Sezione di Pisa, Largo Pontocorvo 3, 56127, Pisa, Italy\\
$^{7}$ INAF-Fundaci\'{o}n Galileo Galilei, Rambla Jos\'{e} Ana Fernandez P\'{e}rez 7, 38712, Bre\~{n}a Baja, Spain\\
$^{8}$ Universit\'{e} C\^{o}te d'Azur, OCA, CNRS, Lagrange, F-06304 Nice, France\\
$^{9}$ INAF-Osservatorio Astronomico di Capodimonte, salita Moiariello 16, 80131, Napoli, Italy\\
$^{10}$ INAF-Osservatorio Astronomico di Bologna, via Ranzani 1, I-40127 Bologna, Italy\\
$^{11}$ INAF-Osservatorio Astronomico di Roma, via Frascati 33, 0040 Monte Porzio Catone, Italy\\
$^{12}$ NSF's National Optical-Infrared Astronomy Research Laboratory, 950 North Cherry Avenue, Tucson, AZ 85719, USA\\
$^{13}$ Department of Physics, Montana State University, P.O. Box 173840, Bozeman, MT 59717-3840, USA\\
$^{14}$ Herzberg Astronomy and Astrophysics, National Research Council, 5071 West Saanich Road, Victoria, British Columbia V9E 2E7, Canada \\
}

\date{Accepted 2019 October 4. Received 2019 September 27; in original form 2019 August 18}

\pubyear{2019}

\begin{document}
\label{firstpage}
\pagerange{\pageref{firstpage}--\pageref{lastpage}}
\maketitle

\begin{abstract}
We present a deep Blanco/DECam colour-magnitude diagram (CMD) for the large but very diffuse 
Milky Way satellite dwarf galaxy Crater~II. The CMD shows only old stars with a clearly bifurcated subgiant branch (SGB) that 
feeds a narrow red giant branch. The horizontal branch (HB) shows many 
RR Lyrae and red HB stars. Comparing the CMD with [Fe/H] = -2.0 and 
[$\alpha$/Fe] = +0.3 alpha-enhanced BaSTI isochrones
indicates a mean age of 12.5 Gyr for the main event and a mean age of 10.5 Gyr 
for the brighter SGB. With such multiple star formation events Crater~II shows similarity to more massive dwarfs that have intermediate age populations, however for Crater~II there was early quenching of the star formation and no intermediate age or younger stars are present. 
The spatial distribution of Crater~II stars overall is elliptical in the plane of the
sky, the detailed distribution shows a lack of strong central concentration, and some inhomogeneities. 
The 10.5 Gyr subgiant and upper main sequence stars show a slightly higher central concentration
when compared to the 12.5 Gyr population. Matching to 
{\it Gaia} DR2 we find the proper motion of Crater~II: $\mu_{\alpha}\cos \delta$=-0.14 $\pm$ 0.07 , $\mu_{\delta}$=-0.10 $\pm$ 0.04 mas yr$^{-1}$, approximately 
perpendicular to the semi-major axis of Crater~II. 
Our results provide constraints on the star formation and chemical enrichment history of 
Crater~II, but cannot definitively determine whether or not
substantial mass has been lost over its lifetime.

\end{abstract}



\section{Introduction}

Crater~II  \citep{torrealba16a} is an enigmatic object in the pantheon of normal and ultra-faint
dwarf galaxies associated with our Galaxy. It occupies a parameter
space where it is physically very large - with half light radius
$1066 \pm 84$ pc \citep{torrealba16a} so similar in size to classic dwarf galaxies such as 
Sculptor and Fornax -
but 100 times less luminous with $M_v = -8.2$ \citep{torrealba16a}, and of such 
low stellar density that its stars are difficult to discern
from amongst Galactic foreground stars and faint background galaxies.  In this latter characteristic it is
similar to the ultra-faint dwarfs (UFD) and \citet{simon19} draws the boundary between UFDs and 
more luminous dwarfs in the luminosity - metallicity plane close to the position of Crater~II.
The discovery of 
Antlia~II \citep{torrealba18} lying behind the Galactic disk and also of similar size
and very low surface brightness might argue for a significant number of such galaxies 
still to be discovered.
It is presently an open question as to how Crater~II formed and subsequently evolved,
with a key aspect being whether it has lost most of its original mass and,
if not, how has it managed to remain intact for a
Hubble time?  \citet{fritz18} analysed the orbits of a number of Milky Way (MW) satellites by
use of {\it Gaia} Data Release 2 ({\it Gaia} DR2, \citealt{gaia18}) proper motion determinations.   
For Crater~II, they find
a reconstructed orbit that is radial with eccentricity $\sim 0.7$ and with pericenter
that may be less than 20 kpc from the center of the MW, depending on the adopted mass for the MW.
Such an orbit, consistent with models by 
\citet{sanders18, fattahi18}, together with the observed size of Crater~II, would suggest a
system vulnerable to disruption. In addition,
\citet{fu19} identify 37 Crater~II members from radial velocities, 22 previously classified
as members by \citet{caldwell17}, and together with 
{\it Gaia} DR2 astrometry calculate 
an orbit for Crater~II, concluding that it is almost certain that the galaxy has been stripped
over its lifetime.  
Spectroscopy of Crater~II members, mostly Red Giant Branch (RGB) and Asymptotic Giant Branch (AGB)
stars \citep{caldwell17, fu19}, demonstrates a resolved but rather narrow range of
[Fe/H] values, suggesting a relatively simple chemical evolution, together
with very cold dynamics although, despite this, apparently Crater~II is dark-matter 
dominated with  M/L~=~53$^{+15}_{-11}$ M$_{\sun}$/L$_{V,\sun}$ \citep{caldwell17}.

Studies of Crater~II 
by our group and others \citep{joo18, monelli18, vivas19}, have identified a copious number of 
RR Lyrae (RRL) variable stars 
indicative of a strong ancient population.  Importantly, RRL are useful for delineating the
spatial distribution of the galaxy, and for large enough structures can determine the depth
as well.   In the case of Crater~II, the RRL distribution could provide a strong observational
constraint as to whether stars are still being lost \citep{vivas19}.   

In this paper we study the star formation history of Crater~II
by analyzing a Colour Magnitude Diagram (CMD) that reaches well below the main sequence
turn-off (MSTO) for the oldest population. A companion paper \citep{vivas19} discusses the 
variable star content of Crater~II, from the same set of observations.
In \S~\ref{sec:data} we describe the
observations and preparation of the data, in \S~\ref{sec:the_cmd} we discuss the
features of the CMD and compare with theoretical isochrones, in \S~\ref{sec:structure}
we discuss the spatial distribution of Crater~II stars, and by matching 80 
stars classified as members and with low {\it Gaia} proper motion errors, we calculate
the Crater~II proper motion. In \S~\ref{sec:variable_stars} we relate the variable stars
properties to the CMD results, and in \S~\ref{sec:discussion}
we discuss the results and present our conclusions.

\section{Observations and Data Reduction}\label{sec:data}

The Crater~II observations presented here derive from an allocation of three
nights awarded by NOAO (P.I. A.R. Walker, prop-id 2017A-0210) with 
DECam on the Blanco 4m telescope at Cerro Tololo Inter-American
Observatory (CTIO).  From these observations we have produced a deep
CMD and 
high-quality light curves for the Crater~II variable stars, with main 
science goals to
study the star formation history and the spatial distribution of the 
galaxy's stellar populations.
The observing log of the observations used for the CMD is summarized in Table~\ref{tab:log}.  
When the Moon was
down we took exposures of 180s in a $u,g,i$ sequence, centered on the Crater~II position
as given by \citet{torrealba16a} with small dithers to fill in the CCD gaps.
If the Moon was up only $g,i$ exposures were taken.  
In photometric conditions SDSS standard star fields were observed, mostly equatorial. 

 \begin{table}
 \begin{center}
 \caption{Log of the DECam observations}\label{tab:log}
 \begin{tabular}{ccccl}
 \hline
 \textit{Run Date} & \textit{$gi$ sequences} & \textit{$ugi$ seq.} & \textit{IQ (arcsec)}  
 & \textit{Comments} \\
  \hline
19 Mar 2017 &    9 & 38 & $0.8-1.5$ & some cloud  \\
20 Mar 2017 &   10 &  44 & $1.0-1.3$ & clear  \\
21 Mar 2017 &   13 & 39 & $1.0-1.5$ & clear  \\
 
 \hline
 \end{tabular}
 \end{center} 
 \label{tab:obslog}
 \end{table}

\subsection{Image Combination}
The images were processed by the DECam Community Pipeline (CP, \citealt{valdes14}), which
removes the instrument signature and provides various data products. For
the specific purposes of this study custom stacks were produced, as
follows:  A total of 153 $g$, 153 $i$ and 32 $u$ band exposures were
available, and after examination of  the image quality and the depth of
each of these exposures, stacks were produced containing 100, 100, 31 selected
images in $g$, $i$, $u$ respectively, rejecting images that were taken in
poor sky transparency, or with a very bright sky, or in poor seeing conditions.   
For these
stacks, the CP produces multi-extension FITS (MEF) files
that have 9 image extensions with each containing approximately 9K $\times$ 9K
pixels and corresponding to $\sim$ 39 $\times$ 39
arcmin on the sky.   The extensions are spatially arranged in a 
3 $\times$ 3 format that covers the  approximately
circular 2 degree diameter DECam field,  thus the four corner extensions
(numbers 1, 3, 7, 9) are only partially filled by DECam.

Two alternative algorithms for flattening the sky background are offered
by the CP.  Here we chose to use the {\it osj} option rather than  
{\it osi} after visual
inspection that the former skies were flatter with only minimal
over-subtraction around very bright objects.  The bad pixel masks provided
for stacks are  very simple, with a value  0 corresponding to good data, a
value 1 coming from an input image bad pixel mask, and a value 2 (the
majority of the bad pixels) coming from where there was no useful data,
e.g. from a star saturated on all images.   Of particular utility is the
exposure map provided by the CP, that is, an image where the data are the
exposure times corresponding to each pixel in the stacked object image.

We further processed the provided stacks in preparation for photometry, producing 
two sets of images as follows: the
first {\it full } set had (i) any pixel appearing with a value of 1 or 2  in the bad
pixel mask was replaced by 32767, the
second {\it uniform } set had (i) any pixel appearing with a value of 1 or 2  in the bad
pixel mask was replaced by 32767, in addition (ii) for all pixels in the exposure maps with
value less than 16200 seconds ($g$, $i$) or 5040 seconds ($u$) the corresponding
image pixel was replaced by 32767.   The values of 16200 and 5040 seconds are 90\% of
the maximum possible exposure times for $g,i$ and $u$ respectively.   The value of 32767 
for the invalid pixel indicator was chosen for compatibility with the 
requirements of the photometry program (see below).

The observation scripts used for the
Crater~II exposures deliberately dithered the telescope just enough to fill in the gaps
between the CCDs, with small additional random dithers of size a few arcsec arising from 
errors in the telescope pointing and adjustments made by the active optics system on an image by image
basis to optimize image quality. The {\it full} data set will therefore contain stars, at a given magnitude, that 
have a range in $S/N$ due to a varying contribution of exposures.  However for brighter 
stars (e.g. including the RRL at g $\sim$ 21) high $S/N$ is achieved 
even with only a few contributing exposures, and the photometric accuracy is dominated 
by systematic effects rather than shot noise.  Therefore, for bright stars we have 
full spatial coverage with no
drawbacks, and we can with confidence use the {\it full} dataset (74541 stars)
for studying the morphology of Crater~II as indicated by the RRL \citep{vivas19}.
For the {\it uniform} set as described above with selection by exposure time,
the sacrifice is that the coverage fraction is reduced to $\sim$ 60 percent (44294 stars)
however the remaining good pixels are very uniform in depth.
This is very important for correct interpretation of any
changes in features in the faintest parts of the CMD as a function of
spatial position. 

In summary, from the above procedures we have produced two sets of three
image FITS files, one each for $g$, $i$ and $u$, each file with 9 image extensions of
approximately 9K $\times$ 9K pixels.  The first set covers the full field with no gaps,
and the second set is more restrictive, including $\sim$ 60 percent of the stars, but with  
uniformity and cleanness suitable for interpreting the 
photometry to the faintest levels.    

\subsection{Photometry}\label{sec:photometry}
We use \verb|DAOPHOT|
\citep{stetson87, stetson94} for the photometry. Firstly, the 
\verb|DAOPHOT| parameter files were prepared, of
particular note is that we set the highest good pixel value ($hi$) = 15000
counts.  While this is much less that the full-well values for any of the
DECam CCDs, the CP does not explicitly correct for the brighter-fatter effect
\citep{bernstein17} so it is preferable to keep stars selected to define the PSF and 
also those used for comparison with photometric standards to
be those of no more than medium brightness.  In any case, the brighter stars 
will all be
Galactic foreground field stars and not of interest to this project.  
We searched for objects down to $S/N$ threshold ($th$) of 3.5,
in two passes.  
For the center 9K $\times$ 9K field (extension 5, centered on Crater~II), and considering
the {\it uniform} sample that excludes the dithered regions, we found 42890,
60368, 1574 objects in $g$, $i$, $u$ respectively thus the object density for $g$, $i$
is approximately one per 90 arcsec$^2$, i.e. comfortably low.

The
Blanco-DECam combination produces tight stellar images and a Moffat function with
beta parameter = 3.5 is an excellent fit to the stellar image profiles, including
the profile wings.  After some
experimentation we chose a linear-varying ($va=1$) PSF as being
appropriate; while a quadratic-varying ($va=2$) PSF showed slightly smaller fit
residuals for the PSF stars, the resulting photometry showed no improvement 
and we decided to keep the simpler functional form for the PSF.   In itself, the 
ability to fit the stacks with a va=1 psf form is a simple confirmation that 
nothing untoward with regard to the star images is taking place spatially 
in the stacking process.

Following aperture photometry and PSF star picking, the PSF photometry was
performed in the standard way with \verb|ALLSTAR|.  For each image the PSF was
constructed from $\sim 200$ stars (100 for $u$).  The photometric errors for
the $g-i$ colors were calculated as the quadrature sum of the errors in $g$
and {i} returned by \verb|ALLSTAR|, and were <0.01 mag for  $g < 21, < 0.02$ mag for $g <
22.5$, and averaged 0.04, 0.1, 0.2 mag at $g = 24,25,26$ respectively.
Since the depths achieved in $g$ and $i$ are similar, the errors in the $g$ and
$i$ bands at these stated magnitudes will be approximately 2/3 of those
for the colors.  Exactly the same procedure was carried out on the 
two sets of images ({\it uniform, full}).

It is instructive to compare the approach chosen here (stack the images, then
run \verb|DAOPHOT/ALLSTAR|), to the alternative of running
\verb|DAOPHOT/ALLSTAR/ALLFRAME| on the individual images.  In principle the
latter forced-photometry approach should be superior \citep{stetson94} 
and be able to go deeper since the star list is
constructed from all available images.   However the computation resources
needed to handle 231 DECam images as opposed to 3 is very significant, and was
beyond those available to our group, even when using \verb|PHOTRED|\footnote{http://github.com/nidever/PHOTRED} \citep{nidever17}
to handle all the book-keeping and efficiently automate the process.   By
contrast, all the photometry for this project was run in a few minutes 
per image using 
a MacBook Pro laptop. The well-known difficulties for doing photometry
on stacks, such as poor control of PSF across the field, poor control of
errors, varying depth, are mitigated here by the low distortion and 
near-constant image quality across the DECam focal plane, by the similariry
of the quantum efficiency response for the DECam CCDs over the passbands of 
main interest ($g$, $i$), and by use of the {\it uniform} set of reductions as 
described above for the critical interpretive tasks.

The photometric ($g$, $i$) calibration involved transformation to the
SDSS system \citep{alam15, nidever17} by cross-correlating 
low-error matches ($\sigma < 0.05$ mag) at the catalogue level
with the Crater~II catalogue produced by \verb|PHOTRED| 
\citep{nidever17}, see \citet{vivas19} for details on how the reference catalogue
was produced. The SDSS and DECam $g$ and $i$ passbands are quite similar, and
we find simple linear transformations, excluding stars with photometric errors
$\sigma>0.05$ mag., and using S and D to denote SDSS and DECam
respectively:
\begin{equation}
g_{S}-g_{D}=0.0710(\pm 0.0003) (g_{S} - i_{S}) + 0.0174(\pm 0.0004),N=7141
\end{equation}
\begin{equation}
i_{S} - i_{D} = 0.0783(\pm 0.0003) (g_{S} - i_{S}) - 0.0541(\pm 0.0004),N=7258
\end{equation}
The $u$ band photometry was left on the DECam natural system, since our intent
was to use these observations to select Crater~II members --mostly RGB, AGB
and Horizontal Branch (HB) stars--
in $g$ vs.  $g-i$ vs. $u-g$ space \citep{dicecco15}.
For this task it is not necessary to calibrate
the $u$ band.

With the photometry in hand, we proceeded with an initial evaluation by
constructing plots of the photometric errors as a function of magnitude,
and plots of the PSF fit parameters {\it chi} and {\it sharp} as functions of
magnitude.   The former allowed formation of an envelope that included the
great majority of detected objects but excluded those with large errors,
and bounds on {\it chi} of $0,1$  fulfilled a similar function of removing errant
measurements. Because of the high Galactic latitude of Crater~II  and
the diffuse nature of the galaxy itself the star density is not so great
as to make photometry difficult by having many overlapping stars, and thus
the {\it sharp} parameter is a very efficient separator of stars and galaxies. 
We used $ -0.5 < sharp < 0.2$ to select stars and $sharp > 0.2$ to select
galaxies.   The samples appear to be very pure (referring here to $g$ and $i$,
the shallower $u$ band images contain few galaxies) except in the final faintest one
magnitude, where from the CMDs (see below) some stars are clearly classified as galaxies, and vice
versa.  With these definitions in
hand, we proceeded to match stars between the photometric bands, taking
account of the systematic few-pixel positional differences between the {$g$, $i$, $u$} stacks,
defining a successful match for objects with centers separated by no more
than 3.5 pixels (0.9 arcsec). In the case of multiple matches, the
closest was selected.  With astrometry from the original images WCS and
redefined using \verb|SAOImageDS9|, the nine photometry files were merged into a single
stellar photometric catalogue.  By comparison, the galaxy photometric catalogue (i.e.
that containing all objects with $ sharp > 0.2$) contains 
approximately four times as many objects as the stellar catalogue.   
The stellar catalogue is shown in Table~\ref{tab:phot}, where we 
use flags to denote whether the object is in the {\it uniform} selection (\textit{F$_{\rm uni}$} = 1), is a variable star (\textit{F$_{\rm var}$} = 1),
or is identified as a probable member on spectroscopic (\textit{F$_{\rm spec}$} = 1) or photometric grounds (\textit{F$_{\rm phot}$} = 1).

\begin{table*}
\centering
\caption{Crater~II Stellar Photometry Catalogue}
\label{tab:phot}
\begin{tabular}{lrrrrrrrrrrr}
\toprule
\textit{ID} & \textit{RA} & \textit{Dec} & \textit{g} & \textit{$\sigma_g$} & \textit{i} &  \textit{$\sigma_i$} & \textit{E(B-V)$_{\rm SFD}$} & \textit{F$_{\rm uni}$*} & \textit{F$_{\rm phot}$*} & \textit{F$_{\rm spec}$*} & \textit{F$_{\rm var}$*} \\
 & (deg) &  (deg) & (mag) & (mag)& (mag) & (mag) & (mag) & & & & \\
\midrule
     1 &  176.16106 & -18.444340 &  19.1575 &    0.003 &  18.4130 &    0.005 &   0.038425 &     0 &     0 &     0 &    0 \\
     2 &  176.16196 & -18.273500 &  25.3828 &    0.119 &  22.9508 &    0.029 &   0.040400 &     0 &     0 &     0 &    0 \\
     3 &  176.16199 & -18.237010 &  24.7683 &    0.091 &  23.7151 &    0.063 &   0.040001 &     0 &     0 &     0 &    0 \\
     4 &  176.16278 & -18.420690 &  21.6401 &    0.008 &  21.0008 &    0.008 &   0.039027 &     0 &     0 &     0 &    0 \\
     5 &  176.16293 & -18.454950 &  25.0343 &    0.087 &  23.5285 &    0.045 &   0.038168 &     0 &     0 &     0 &    0 \\
     6 &  176.16348 & -18.429810 &  25.6254 &    0.162 &  24.5563 &    0.115 &   0.038752 &     0 &     0 &     0 &    0 \\
     7 &  176.16354 & -18.247700 &  23.4218 &    0.020 &  21.8672 &    0.012 &   0.040073 &     0 &     0 &     0 &    0 \\
     8 &  176.16368 & -18.262930 &  22.2587 &    0.008 &  20.5067 &    0.005 &   0.040219 &     0 &     0 &     0 &    0 \\
     9 &  176.16371 & -18.438970 &  20.8335 &    0.004 &  20.3301 &    0.006 &   0.038505 &     0 &     0 &     0 &    0 \\
    10 &  176.16373 & -18.227880 &  24.9655 &    0.087 &  21.8863 &    0.010 &   0.039807 &     0 &     0 &     0 &    0 \\
    ... & ... & ... & ... & ... & ... & ... & ... & ... & ... & ... & ... \\
 37914 &  177.32239 & -18.648880 &  21.0050 &    0.039 &  20.7230 &    0.017 &   0.033798 &     0 &     0 &     0 &    1 \\
 37915 &  177.32240 & -17.783121 &  17.8570 &    0.002 &  17.3552 &    0.002 &   0.030800 &     1 &     0 &     0 &    0 \\
 37916 &  177.32245 & -17.446171 &  25.6924 &    0.091 &  24.2554 &    0.062 &   0.035800 &     1 &     0 &     0 &    0 \\
 37917 &  177.32245 & -18.545030 &  24.5695 &    0.049 &  24.2915 &    0.076 &   0.032083 &     0 &     0 &     0 &    0 \\
 37918 &  177.32253 & -18.674610 &  25.0724 &    0.094 &  24.7120 &    0.122 &   0.033977 &     0 &     0 &     0 &    0 \\
 37919 &  177.32253 & -18.654630 &  24.0953 &    0.037 &  23.2545 &    0.036 &   0.033842 &     0 &     0 &     0 &    0 \\
 37920 &  177.32254 & -18.543490 &  25.7842 &    0.123 &  23.5299 &    0.034 &   0.032056 &     0 &     0 &     0 &    0 \\
 37921 &  177.32256 & -18.089280 &  25.3239 &    0.088 &  24.6449 &    0.106 &   0.033353 &     0 &     0 &     0 &    0 \\
 37922 &  177.32257 & -18.374910 &  24.2701 &    0.030 &  23.9927 &    0.054 &   0.030500 &     1 &     0 &     0 &    0 \\
 37923 &  177.32257 & -18.273441 &  23.7604 &    0.020 &  23.7887 &    0.048 &   0.033800 &     1 &     0 &     0 &    0 \\
 37924 &  177.32257 & -18.932541 &  25.8975 &    0.117 &  25.3470 &    0.173 &   0.038400 &     1 &     0 &     0 &    0 \\
 37925 &  177.32262 & -18.363504 &  24.0402 &    0.025 &  21.8884 &    0.007 &   0.030500 &     1 &     0 &     0 &    0 \\
 37926 &  177.32269 & -18.632470 &  19.5170 &    0.003 &  18.3982 &    0.003 &   0.033608 &     0 &     1 &     1 &    0 \\
 37927 &  177.32269 & -18.947861 &  25.5766 &    0.090 &  24.9110 &    0.121 &   0.038900 &     1 &     0 &     0 &    0 \\
 37928 &  177.32269 & -18.797030 &  19.6891 &    0.003 &  18.2289 &    0.003 &   0.036104 &     0 &     0 &     0 &    0 \\ 
 ... & ... & ... & ... & ... & ... & ... & ... & ... & ... & ... & ... \\
 74532 &  178.45450 & -18.356200 &  23.5320 &    0.025 &  23.0068 &    0.028 &   0.037385 &     0 &     0 &     0 &    0 \\
 74533 &  178.45505 & -18.378880 &  25.3801 &    0.109 &  24.3159 &    0.083 &   0.037335 &     0 &     0 &     0 &    0 \\
 74534 &  178.45630 & -18.438780 &  22.4640 &    0.011 &  20.9631 &    0.011 &   0.039191 &     0 &     0 &     0 &    0 \\
 74535 &  178.45670 & -18.398970 &  24.5338 &    0.065 &  22.3133 &    0.015 &   0.038049 &     0 &     0 &     0 &    0 \\
 74536 &  178.45689 & -18.326450 &  21.8885 &    0.008 &  19.4466 &    0.004 &   0.037705 &     0 &     0 &     0 &    0 \\
 74537 &  178.45721 & -18.385570 &  23.5345 &    0.027 &  22.8753 &    0.025 &   0.037589 &     0 &     0 &     0 &    0 \\
 74538 &  178.45842 & -18.466220 &  24.1100 &    0.039 &  21.2650 &    0.009 &   0.039506 &     0 &     0 &     0 &    0 \\
 74539 &  178.45905 & -18.396870 &  25.5048 &    0.117 &  24.2044 &    0.084 &   0.038045 &     0 &     0 &     0 &    0 \\
 74540 &  178.45979 & -18.365350 &  24.4644 &    0.053 &  21.8250 &    0.012 &   0.037323 &     0 &     0 &     0 &    0 \\
 74541 &  178.46194 & -18.465880 &  24.5687 &    0.069 &  22.5059 &    0.020 &   0.039691 &     0 &     0 &     0 &    0 \\
\bottomrule
\end{tabular}
\begin{tablenotes}
\item *\textit{F$_{\rm uni}$}, \textit{F$_{\rm phot}$}, \textit{F$_{\rm spec}$}, and \textit{F$_{\rm var}$} are flags that show if a particular star has been identified in the uniform catalogue, as photometric member, as spectroscopically-confirmed star, and as variable star member, respectively.
\item \textit{Notes.-} Table~\ref{tab:phot} is published in its entirety in the machine-readable format. A portion is shown here for guidance regarding its form and content.
\end{tablenotes}
\end{table*}

\section{The Colour Magnitude Diagram}\label{sec:the_cmd}

\subsection{Reddening Correction}\label{sec:reddening}

\begin{figure}
\vspace{-0.5cm}
\hspace{-1cm}
\includegraphics[width=0.5\textwidth]{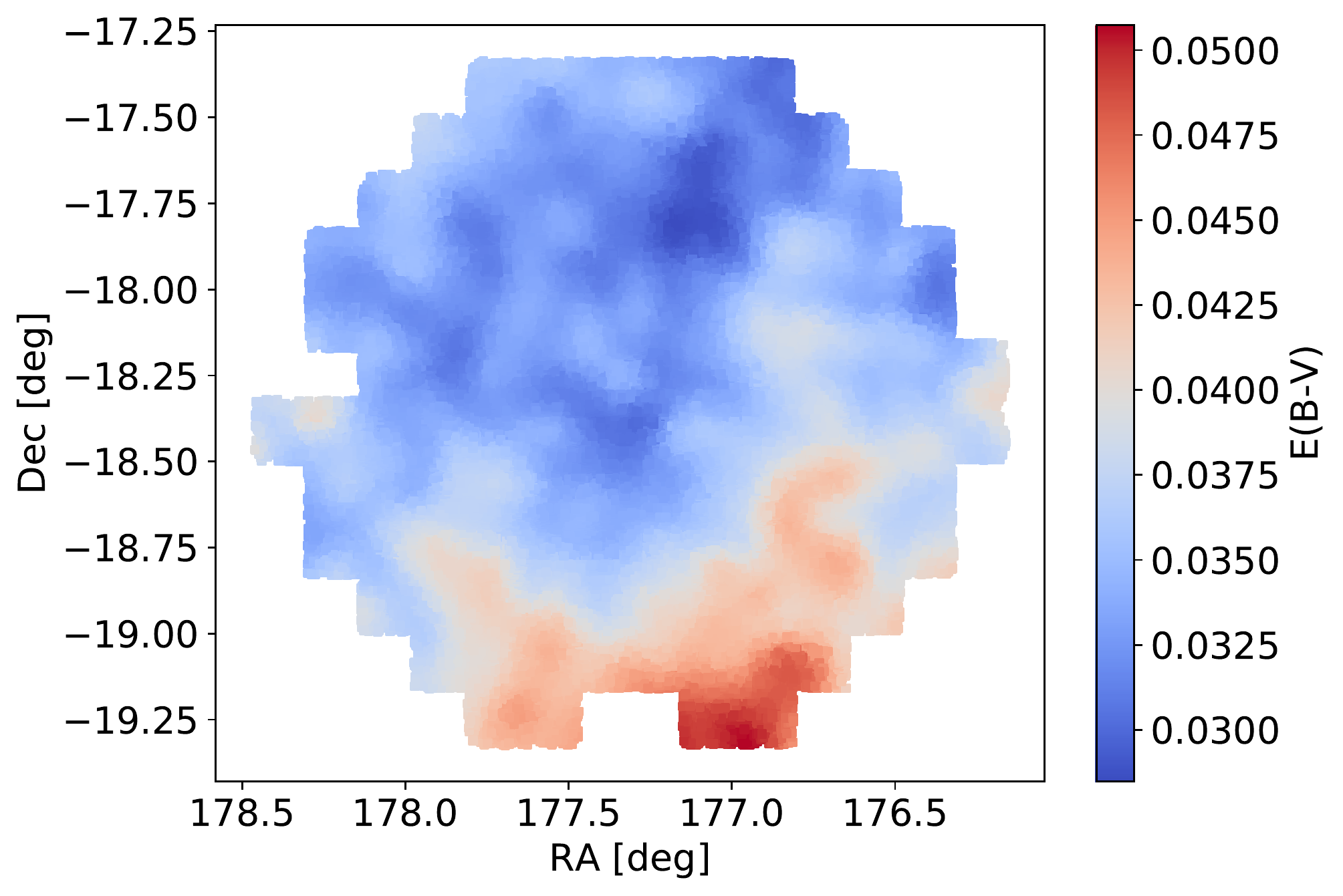}
\caption{Interstellar dust reddening map from \citet{schlegel98} of the central $\sim$3 sq. deg. of Crater~II observed with DECam.}
\label{fig:reddening_map}
\end{figure}

We first correct the Crater~II photometry for foreground reddening, obtaining E($B$-$V$) from the \citet{schlegel98} interstellar dust maps using the python task \verb|dustmaps| \citep{green18}. Although the reddening is small due to the high Galactic latitude, the very large field of DECam usually means that the reddening varies across the field, and so this aspect should be taken into account (see Figure~\ref{fig:reddening_map}).  
In order to apply the reddening correction, we correct each $g$ and $i$ magnitude of our detected sources using the \citet{schlegel98} reddening values and the coefficients given in \citet{schlafly11}, i.e. A$_g$ = 3.303 E($B$-$V$), A$_i$ = 1.698 E($B$-$V$). The same procedure was applied in the companion paper \citep{vivas19}.

The reddening-corrected $i, g-i$ CMDs are shown in Figure~\ref{fig:cmd}, where results for both the {\it full} and {\it uniform } reductions are displayed, each of the two sets of three panels depict stars within a elliptical distance\footnote{We use the elliptical geometry for Crater~II as determined from the RRL \citep{vivas19} so the elliptical distance stated is half of the geometrical constant of the ellipse at the position of each star.  We denote this by using the nomenclature r'.} of 0.35 degrees of the center of Crater~II, between 0.35 and 0.65 degrees, and exterior to 0.65 degrees. We denote these inner, central, and outer regions, respectively.

\subsection{CMD Description}\label{sec:cmd_description}

The CMD (Figure~\ref{fig:cmd}) reaches to $ g,i \sim 25.5$ and 
the stellar sequences belonging to Crater~II are thus defined to well below the level
of the oldest MSTO.  Crater~II stars are prominent in the inner and central regions,
while the outer region is dominated by field stars.   In the outer region of
the DECam field,  compact and relatively blue galaxies dominate in numbers over both field and Crater~II
stars for magnitudes fainter than $i \sim 24.5$.  With  low $S/N$ some galaxies will have measured
value of \verb|DAOPHOT| $sharp < 0.2$ and will thus be classified as stars, and given the relative numbers,
will appear on the CMD.
It is notable that this contamination is less prominent in the {\it uniform} sample than for the 
{\it full} sample, as expected since the {\it uniform} sample will exclude stars with low $S/N$ for a given
magnitude.
Over the whole field, compact galaxies outnumber stars by a factor
three, but as previously discussed a cut on \verb|DAOPHOT| {\it sharp} excludes these with high efficiency except near the faint 
magnitude cutoff.

Crater~II shows a strongly populated HB in the vicinity of the RRL, the latter are
relatively easy to see on the CMD since they are bluer than the majority of field stars. There are
red HB (RHB) stars, much more contaminated by MSTO field stars than the RRL. There is a likely increase 
in density of the RHB stars of stars near the red end of the RHB.  There appear to be no blue HB (BHB) stars
at all, and this
is consistent with the RRL distribution, see discussion in \citet{vivas19}.  The RGB is
narrow, down to the base, however the subgiant branch (SGB) is clearly split
into two components, both feeding into the RGB. There is an indication that the 
brighter SGB is less prominent at larger radii, this will be discussed in the following section.
There are many blue stragglers, but no turn-off brighter than the two just described is visible.

The Crater~II AGB stars and RGB stars brighter than the HB are difficult to discern 
from amongst the foreground field stars, even in the inner regions.  However there are three 
 ways in principle that we can identify Crater~II stars in the field-star dominated regions of the CMD.
 
Firstly, we can cross reference our photometry to stars classified as members by 
\citet{caldwell17,fu19} on the basis of radial velocity and metallicity; we call this the
\textit{spectroscopic} membership sample. There are a total of 70 stars in this sample, 56 of them are 
in \citet{caldwell17} and 35 in \citet{fu19}, with 21 stars in common between both catalogues. It is worth
noting that, by definition, this is a very pure sample.

\begin{figure*}
\includegraphics[width=0.43\textwidth]{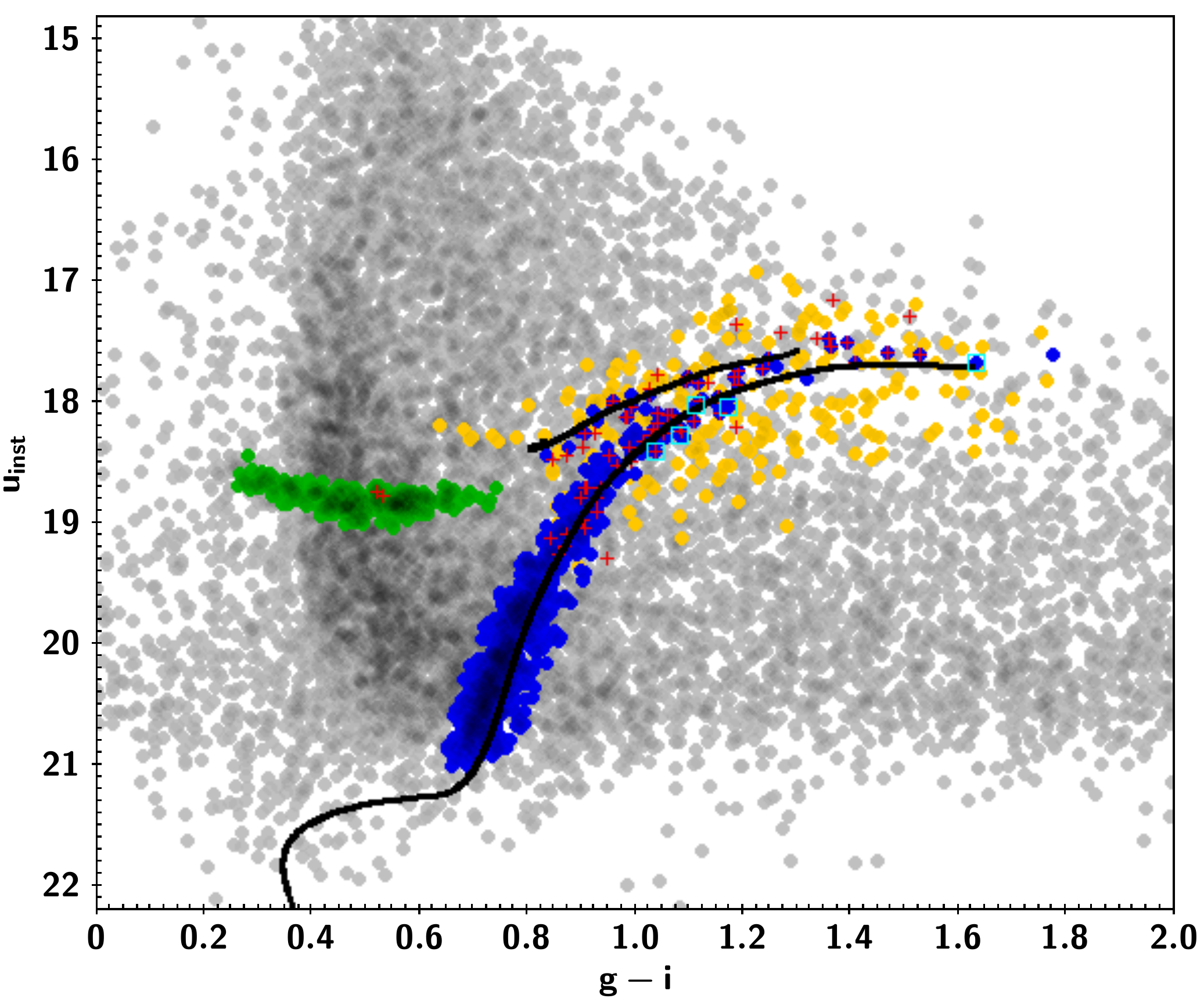}
\includegraphics[width=0.55\textwidth]{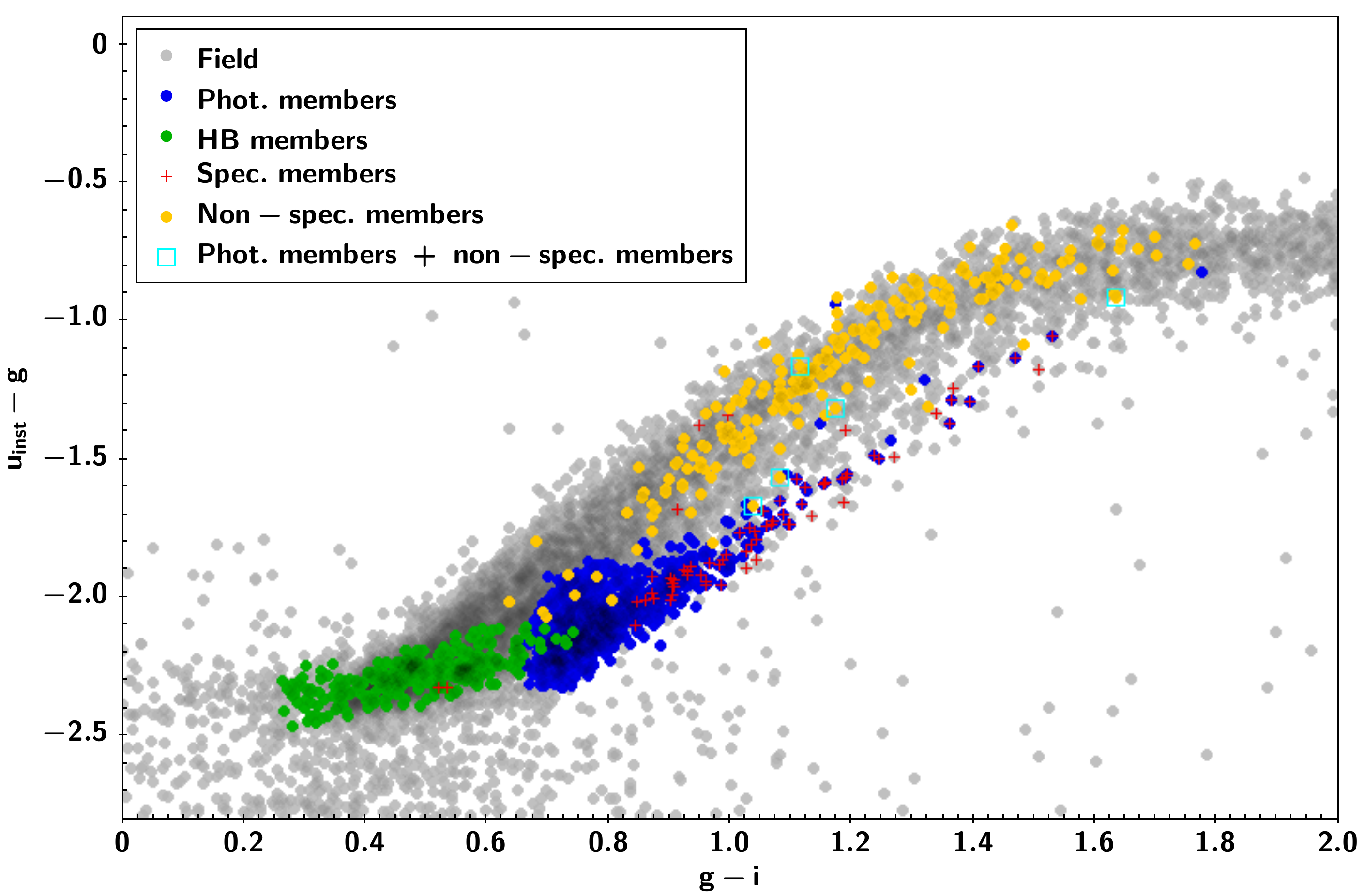}
\caption{Left.- ($u_{\rm inst}$, $g-i$) colour-magnitude diagram of the sources contained within the DECam FoV centered on Crater~II. Field stars are shown in grey, photometric members in blue (AGB+RGB) and in green (HB), spectroscopically confirmed members as red crosses and spectroscopically confirmed non-members in yellow. Cyan open squares display those stars that were classified in this work as photometric members of Crater~II but they are not members from the spectroscopy. A 12.5 Gyr isochrone of [Fe/H] = -2.0 dex is plotted for guidance. Right.- Colour-colour diagram for the same sources. The spectroscopically confirmed non-members are located in the region of the more metal-rich stars, while the members of Crater~II are in the region of the most metal-poor stars.}
\label{fig:color_color}
\end{figure*}

Secondly, we have $u$ band photometry for all stars to below the level of the HB, and so can select for stars that are metal-poor in the two-colour diagram, and also lie close to the cluster sequences in the CMD. In order to separate candidates and field stars, we followed the prescription described by \citet{dicecco15} and \citet{calamida17}. Briefly, we generated two isodensity maps in the $i$, $u-g$ and in the $g, u-g$ CMDs and created two ridge lines that helped us to locate our most probable candidates. These candidates were selected considering those stars located between $\pm$1$\sigma$ (defined as the quadratic sum of the photometric errors in the three bands) to the 3D ridge line and their position in the ($u-g$), ($g-i$) diagram. We selected a total of 1166 stars that we call the \textit{photometric} catalogue. This method is expected to work well for AGB and RGB stars because the magnitude of these stars in the $g$ and $i$ bands is a strong function of the colour, and thus provides an effective constraint.  We can select HB stars similarly, although the small range in luminosity makes field star discrimination less effective. Figure~\ref{fig:color_color} shows, in the left panel, the ($u_{\rm inst}$, $g-i$) CMD and, in the right panel, the ($u_{\rm inst} - g$) versus ($g-i$) colour-colour diagram for the selected photometric membership sample with blue (RGB+AGB) and green (HB) dots while red crosses represent the spectroscopic membership sample.

We can test for field star contamination in the RGB and AGB photometric catalogue (863 stars) in the following way: \citet{caldwell17} in their table 2 
list all their observed stars, and have provided probabilities (private communication to M. Monelli) that each star is a Crater~II member.  All except a very few stars are classified as member (probability 1) or non-member (probability 0), see \citet{caldwell17} for details of how they calculate the probability.  There are 314 stars with probability 0, represented as yellow dots in Figure~\ref{fig:color_color}. We cut off our photometric catalogue at $i = 20.1$ to match the spectroscopic catalogue, leaving 157 stars,
and search for positional 
matches; we find five (cyan open squares in Figure~\ref{fig:color_color}) stars with positional agreements 0.2-0.3 arcsec using a 5 arcsec window.  With the areas of the
two catalogues differing by a factor 2.0, then the contamination in our catalogue of RGB and AGB stars is thus 5 $\times$ 2 / 157 = 0.06.  This analysis does assume that the fainter stars in our photometric catalogue have the same degree of
field star contamination as do those brighter than $i = 20.1$ but this is a reasonable assumption, see Figure~\ref{fig:color_color}.  We note that one
of the five stars is metal poor, [Fe/H] = -1.67 and with a radial velocity indicating a non-member, is apparently a halo giant at approximately the distance of Crater~II.  The remaining four stars are non members on the grounds of both metallicity and radial velocity.  

The photometric sample also contains 303 RHB stars.  Unfortunately, we have no 
way of testing spectroscopically for membership in any significant way, with only two RHB stars in the \citet{fu19} catalogue.
Membership can be evaluated by counting field stars in photometric (magnitude, colour) boxes above and below the RHB.  This is complicated by the high density of field stars, and the rapid change of star density as a function of colour as the MSTO colour for the foreground stars is approached. We conclude that the photometric catalogue for the RGB and AGB stars is pure at the 95\% level, but the RHB star entries (303 stars) will be contaminated by field
stars at a much higher level, and one that is difficult to estimate.  

Thirdly, we have the variable star members. In our companion paper \citep{vivas19}, we have identified a total of 
106 variable star members of Crater~II which consist of 98\footnote{There are 99 RRL members of Crater~II but one, 
discovered by \citet{joo18} is outside our DECam field of view.} RRL, seven Anomalous Cepheid (AC), and one dwarf 
Cepheid (DC) stars. This sample is also considered a very pure.

\begin{figure*}
\includegraphics[width=1.0\textwidth]{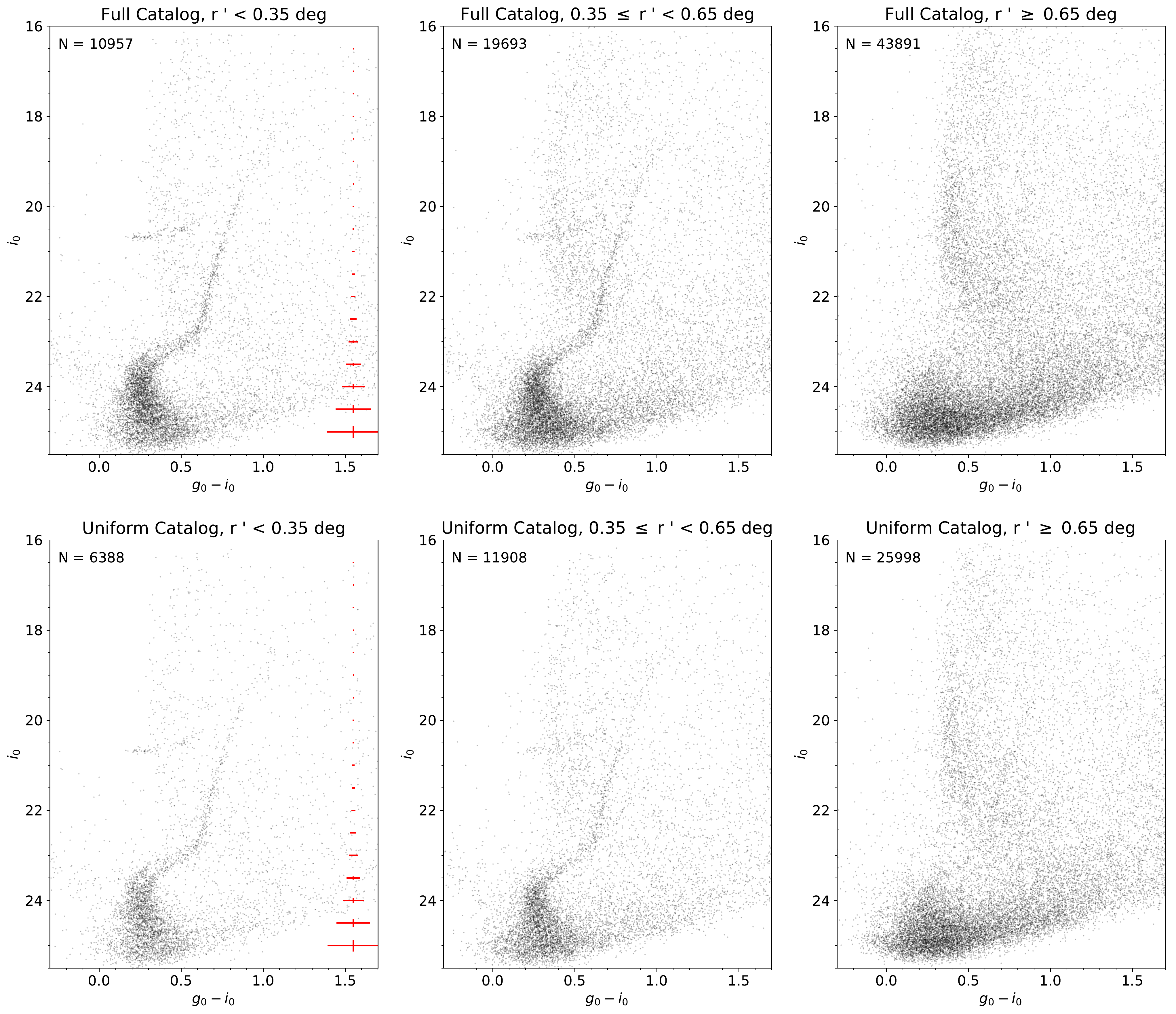}
\caption{Colour-Magnitude diagrams of Crater II. Top and bottom panels represent the full and the uniform catalogues, respectively (see the text for details). Left panels show those stars within 0.35\degr elliptical distance (see text for details) centred on Crater~II,  middle panels display those stars within 0.35\degr and 0.65\degr, while right panels show those stars beyond 0.65\degr.}
\label{fig:cmd}
\end{figure*}

\subsection{Isochrone Comparison to the CMD}\label{sec:isochrone_fitting}

\begin{figure*}
\includegraphics[width=1.0\textwidth]{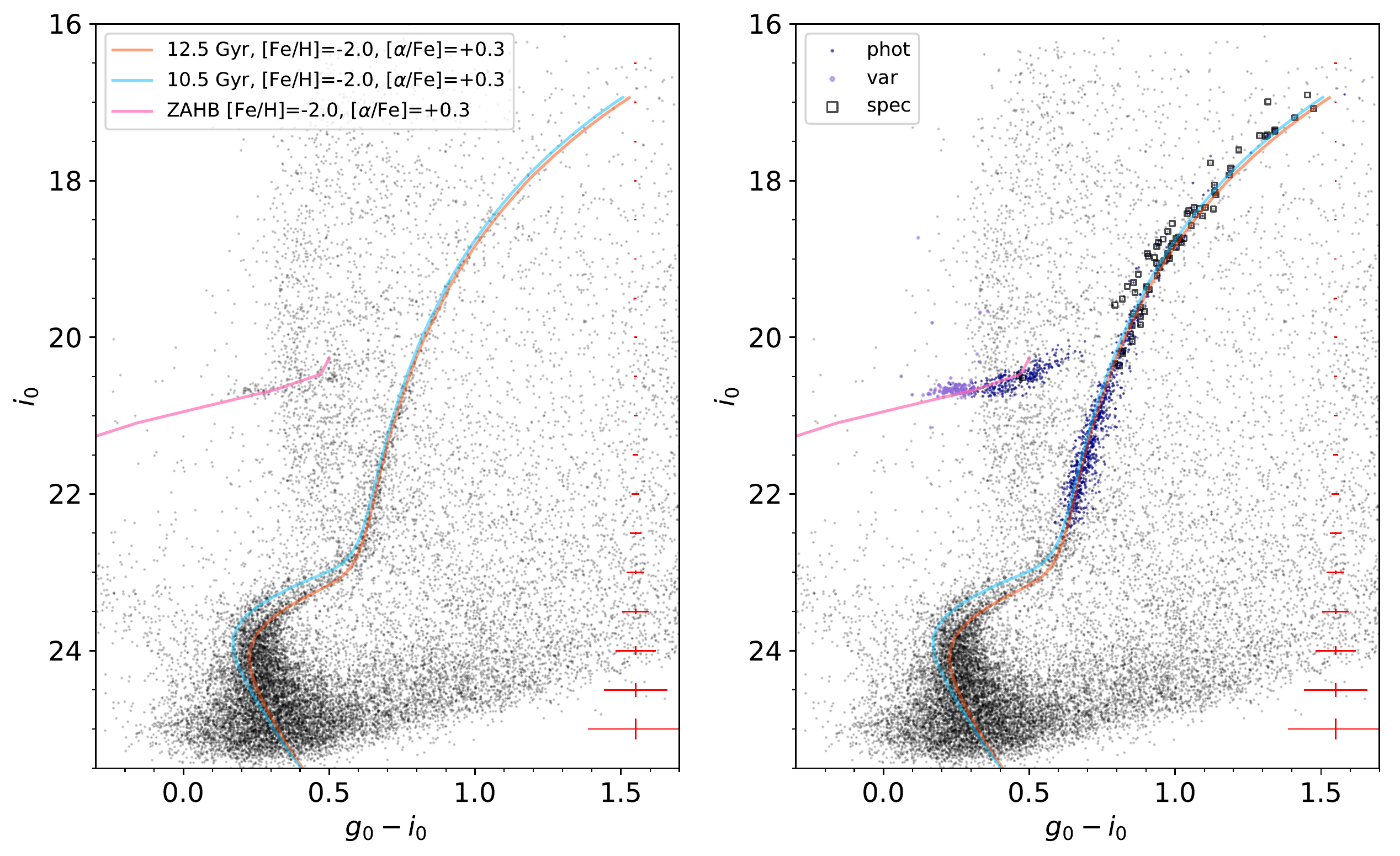}
\caption{Isochrone Comparison to the CMD of Crater II (r' $<$ 0.78\degr, 1.5$\times$r$_h$). The CMD was corrected star-by-star by the \citet{schlegel98} reddening map, using the modified version of \citet{schlafly11}. Orange and light blue lines represent the best set of [Fe/H]=-2.0 alpha enhanced ([$\alpha$/Fe]=+0.3) isochrones from BaSTI (Pietrinferni et al. in preparation) that match the CMD of Crater II, 12.5 Gyr and 10.5 Gyr respectively.  Isochrones and ZAHB were shifted using a distance modulus for Crater~II of 20.33 mag \citep{vivas19}. Additionally, we add 0.04 mag to the total reddening in order to make the isochrones and the stellar sequences match better (see text for more details). Spectroscopic members are shown as black squares in the right panel, while photometric members are displayed as dark blue dots and variable members (RR Lyrae stars, Anomalous Cepheids, dwarf Cepheid) are purple dots.}
\label{fig:isochrones}
\end{figure*}

The BaSTI\footnote{http://basti-iac.oa-abruzzo.inaf.it} \citep{hidalgo18} model grid selected for the present analysis corresponds to that of stellar evolutionary computations accounting for the occurrence of mass loss (according to the Reimers’ law and the free parameter $\eta$ set to the value of 0.3) as well as for core convective overshooting during the central H-burning stage and atomic diffusion.  However core convective overshooting is irrelevant in the present context since low-mass (i.e. old) stars  burn H into a radiative core.  In Figure~\ref{fig:isochrones} we compare the CMD sequences with 
alpha-enhanced (Pietrinferni et al. in prep.) BaSTI isochrones and zero-age horizontal branch (ZAHB) models.

The spectroscopy by \citet{caldwell17,fu19} finds mean [Fe/H] = -2.0 with a small dispersion (sigma $\sim 0.2 \pm 0.1$~dex). With Crater~II lying near the boundary between UFDs and classical dwarfs \citep{simon19}, we assume that Crater~II will behave similarly with respect to alpha element enhancement. All these galaxies, at the metallicity of Crater~II or lower, are alpha-enhanced \citep{kirby11,vargas13,simon19} with the transition from high alpha to solar alpha ratio driven by the change from dominance of SN II to SN Ia, and showing variance from galaxy to galaxy depending on the details of the star formation history for each.  For Crater~II a choice of [$\alpha$/Fe] = +0.3 would seem appropriate. We therefore choose BaSTI-IAC alpha-enhanced isochrones (Pietrinferni et al. in preparation) with [Fe/H] = -2.0 and [$\alpha$/Fe] = +0.3, after exploring parameter space quite widely.
In matching the isochrones and the ZAHB to the data we use the most recent RRL distance modulus
($\mu_0$ = 20.333 $\pm$ 0.004 mag, \citealt{vivas19}), and apply a small colour shift of 0.04 magnitude 
so that the isochrones better coincide with the 
observational sequences. The origin of this colour shift could be one or more of small errors in the 
reddening scale, errors in the photometric zeropoints, or 
 a residual shortcoming in the adopted colour-effective temperature scale adopted for transferring the 
 isochrones from the theoretical plane to the observational one.

The right panel of Figure~\ref{fig:isochrones} shows that the photometric (dark blue dots) and spectroscopic 
(black open squares) members of Crater~II play an important role when doing the isochrone matching.
The age of the
isochrone that best matches the older turnoff is 12.5 Gyr and the younger SGB is best matched with an 
isochrone with the same metallicity and an age of 10.5 Gyr.  The latter is a minority population,
and given the errors in the individual spectroscopic measurements for the stars classed as 
Crater~II members, we cannot rule out that the 10.5 Gyr stars might have slightly lower alpha elements 
abundance and slightly higher [Fe/H] than does the 
12.5 Gyr population, a scenario similar to that convincingly demonstrated for the Carina dwarf  
galaxy by \citet{vandenberg15}.   If, for example, we compare a solar-scaled isochrone with [Fe/H] = -1.7 
\citep{hidalgo18} to the younger population then an excellent match is obtained, although at an age 1 Gyr younger (9.5 Gyr)
than the alpha enhanced isochrone.

Similarly, while there is no obvious evidence for older (12.5 - 13.5 Gyr) stars,
the lack of sensitivity of the RGB colour to metallicity for metal poor stars, the age-metallicity 
degeneracy, and the unknown alpha element enhancement for such stars, make it possible to hide 
such a minority population.  We return to this issue in \S~\ref{sec:discussion}.

Note that the isochrones are placed on the centre of the star distributions in the CMD; compared to 
the magnitude errors at the level of the SGBs both appear to be broadened in age. 
The younger population is close to, or under, the age limit of 
10-11 Gyr for forming RRL \citep{walker89, glatt08, catelan18},
and is thus expected to contain few or no RRL, with the majority of the
core-helium burning stars populating the RHB.

\subsection{The Horizontal Branch Morphology}\label{sec:hb_morph}

We consider an inner region with r' $<$ 35 arcmin (see \S~\ref{sec:spatial_distribution}) within which we count 0 BHB, 43 RRL and 80 RHB stars, and (see Table~\ref{tab:rad_dist}) 530 stars from the older SGB and 233 from the younger. The older SGB will populate the RHB and RRL, while the younger SGB will likely only populate the RHB, if so we calculate that 43 RRL and 42 RHB stars belong to the 12.5 Gyr population and 38 RHB stars belong to the 10.5 Gyr population. We calculate the HB morphology parameter \citep{lee90} for the 12.5 Gyr population to be $\frac{\rm{N(BHB)} - \rm{N(RHB)}}{\rm{N(BHB)} + \rm{N(RRL)} + \rm{N(RHB)}} = -0.49$.

As demonstrated by
\citet{lee94} and  \citet{sara95} for Galactic globular clusters, a HB dominated by RRL and 
RHB stars with few or no BHB stars can be produced by high metal abundance, clearly not the case 
here, but also by young age.  For a metal poor population such as Crater~II, an age of the RRL of 11-12 
Gyr after applying a small zeropoint offset, accounting for the modern age of the Universe (13.8 Gyr), to
the results of \citet{lee94} and \citet{sara95}, is consistent with the observed HB morphology, but
suggesting an age slightly younger than found by our isochrone fitting.

Other parameters that can affect the HB morphology are He abundance enhancements or spread, or a non-canonical value for the mass loss on the RGB. After the discovery of multiple stellar populations in globular clusters there is compelling evidence \citep{milone14} that, after metallicity and age, He abundance anomalies play a major role in determining HB morphology. However for Crater~II such anomalies would produce BHB stars, which are not observed. For mass loss, \citet{mcdonald15} show for a large sample of globular clusters that there is little dispersion in the measurements. We will proceed for Crater II by making the assumption that age and metallicity are driving the HB morphology, as there is no evidence from the available data that more exotic explanations are required.

\subsection{The CMD and the RR Lyrae Variables}\label{sec:cmd_rrl}

\citet{joo18,monelli18}; \citet{vivas19} show that  the Crater~II RRL period distribution is unusual with only a few 
 RRcd stars and many RRab. This distribution implies that a 12.5 Gyr stellar population, during 
 the core He-burning stage, is not able to populate the whole RR Lyrae domain from the
 blue edge of the First Overtone instability strip to the red boundary of the Fundamental  strip.
 In this respect Crater~II is similar to the unusual Galactic
globular cluster Ruprecht 106
\citep{dotter18}, which is young, metal-poor and contains only RRab 
variables \citep{buonanno93,kaluzny95,leaman13}.

By considering the dispersion in brightness in the $i$-band RRL period-luminosity (PL) relation, 
and dividing the RRab stars into bright (B) and faint (F) groups by considering the
position of each star relative to the mean PL relation,
\citet{vivas19} 
show that the two groups have a different spatial distribution with the F group more 
centrally concentrated (their figure 12).  This is interpreted as a small difference in 
the mean metallicity (nominally 0.17 dex) between the two groups. 
We do note, considering the spectroscopic sample of stars
and removing the AGB stars, the RGB stars are almost all closer than 0.02 mag in colour
to the fitted isochrone (Figure~\ref{fig:isochrones}).  The BaSTI isochrones show that a small 
systematic metallicity shift 
of 0.1-0.2 dex could be hidden in the present observations particularly if combined with a 
slight age change (e.g. a few 100 Myr) given the  well-known
degeneracy between RGB age and metallicity, in the sense that younger age will produce a bluer 
RGB and higher metallicity will make the RGB redder. 
   
Additionally, we plotted a C$_{ugi}$ diagram
\citep{monelli13} for the Crater~II RGB stars with spectroscopy, this pseudo-colour can split stars with 
different metalliicities over much of the RGB. We divided the stars 
into two groups divided by [Fe/H] = -2.0, and found no significant difference between the 
location of the two groups of stars in this diagram.   We conclude that we are not able, 
with the photometry alone, to provide further insight into the possible metallicity gradients in 
Crater~II suggested by \citet{vivas19}.


\section{Crater~II Structure and Motion}\label{sec:structure}

\subsection{Proper Motion}\label{sec:proper_motion}

\begin{figure*}
\includegraphics[width=0.99\textwidth]{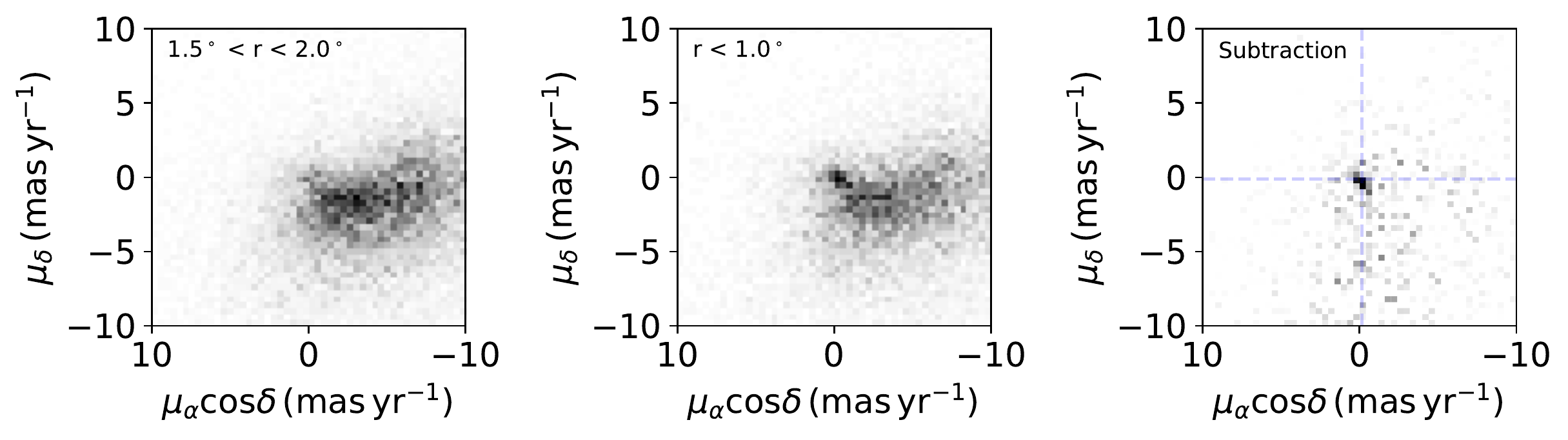}
\caption{Left.- Proper motions of a field region defined by a circular ring of 1.5\degr $<$ r $<$ 2.0\degr centered on Crater~II. Middle.- Proper motions of a circular area inside r $<$ 1.0\degr centered in Crater~II. Right.- The subtraction between left and middle panel reveal the proper motion of Crater~II. The intersection of the two dashed blue lines marks the locus of the proper motion determined by the members of Crater~II (see text for details).}
\label{fig:pm_field}
\end{figure*}

\begin{figure}
\hspace{-1cm}
\includegraphics[width=0.48\textwidth]{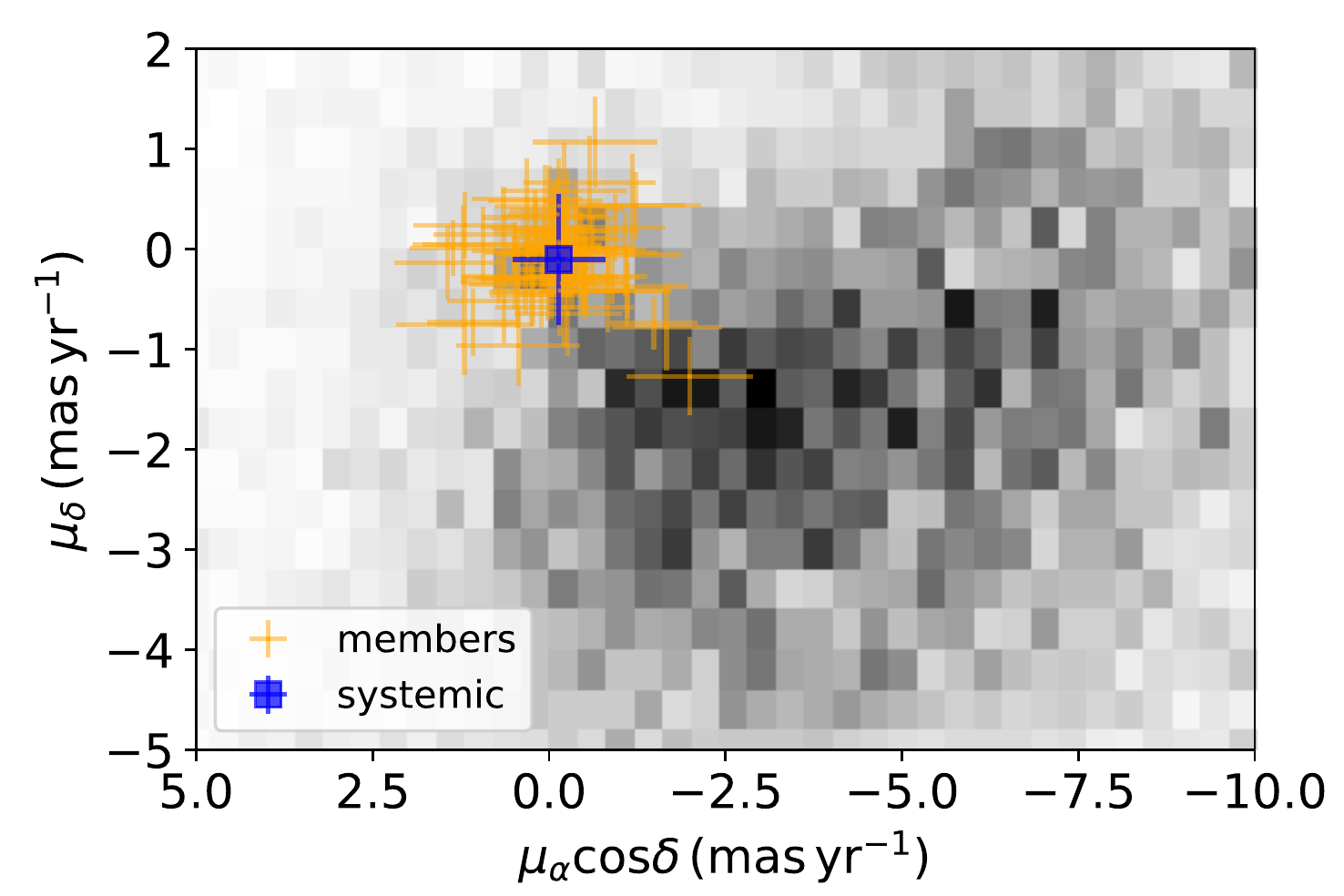}
\caption{Proper motions of the Crater II photometric, spectroscopic, and variable members (orange) present in our catalogue whose proper motion errors are smaller than 1 mas yr$^{-1}$. The blue square shows the 
systemic proper motions of Crater II obtained from the selected (80) stars: $\mu_{\alpha} \cos{\delta}$ = -0.14 $\pm$ 0.07 mas yr$^{-1}$, 
$\mu_{\delta}$ = -0.10 $\pm$ 0.04 mas yr$^{-1}$. The grey scale map shows the proper motions of a field mostly
outside Crater~II, defined by a region of 1.5\degr $<$ r $<$ 2.0\degr centered on Crater~II. }

\label{fig:pm}
\end{figure}

 All proper motion measurements are from {\it Gaia} DR2 \citep{gaia18}. Figure \ref{fig:pm_field} has three panels, the 
 left panel shows the 2D proper motion distribution for stars in an external field comprising
 an annulus with radii of 1.5 and 2.0 deg centered on Crater II, the center panel
 shows the 2D proper motions distribution for stars inside r < 1.0 deg, and the right panel is the subtraction 
 between the first two panels (normalized) that clearly reveals Crater~II. Guided by this,
 we proceed by selecting stars in our catalogue that satisfy the following criteria: they are
 flagged as photometric, spectroscopic, or variable star members; they are found in {\it Gaia} DR2 with proper 
 motion errors smaller than 1 mas yr$^{-1}$; and they have proper motions between $\pm 3$ mas yr$^{-1}$ mas yr$^{-1}$ 
 (both in RA and Dec).  This gives a sample of 80 stars, which are displayed as orange symbols in Figure~\ref{fig:pm}, 
 of which 50 have spectroscopy and thus also appear in proper motion analyses by \citet{fritz18,fu19}, 65 are photometric 
 members (of which 36 have spectroscopy), and 1 AC.  The proper motion of Crater~II is then determined by a weighted 
 average of the proper motions for these 80 stars, in units of mas yr$^{-1}$,
$\mu_{\alpha} \cos{\delta} = -0.14$  $\pm$ 0.07 (standard deviation = 0.66), 
$\mu_{\delta} = -0.10$ $\pm 0.04$ (standard deviation = 0.38), represented by a blue square in Figure~\ref{fig:pm}. This Crater~II systemic proper motion, using a slightly larger sample of stars, is 
consistent within the errors with the two previous {\it Gaia} derived proper motions 
(\citealt{fritz18}: $\mu_{\alpha} \cos{\delta}$ = -0.18 $\pm$ 0.06, $\mu_{\delta}$ = -0.11 $\pm$ 0.03 
from 58 \citealt{caldwell17} spectroscopic members; 
\citealt{fu19}: $\mu_{\alpha} \cos{\delta}$ = -0.17 $\pm$ 0.06, $\mu_{\delta}$ = -0.07 $\pm$ 0.07
from 37 of their spectroscopic members).

\subsection{The Spatial Distribution of Crater~II stars}\label{sec:spatial_distribution}

\begin{figure}
\includegraphics[width=0.48\textwidth]{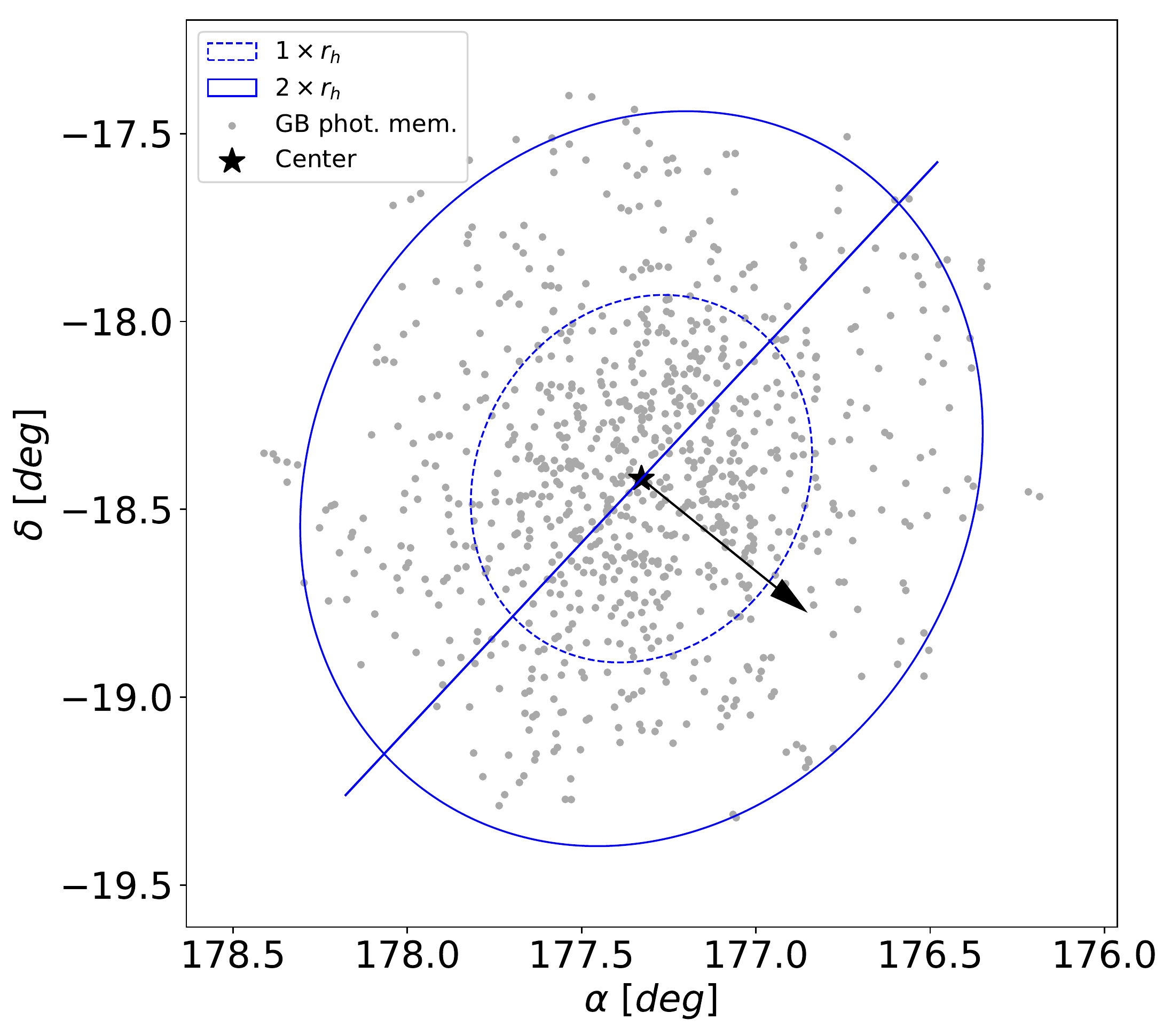}
\caption{Spatial distribution of the giant branch (RGB + AGB) stars considered photometric members (grey dots) of Crater II. Morphological parameters were calculated using a robust bivariate normal distribution fit to these stars. The ellipticity (0.12) and position angle (135\degr) of the blue ellipses located at 1 and 2$\times$ r$_h$ \citep{torrealba16a} come from the results of this fit. The new recalculated center of Crater II is marked with a black star. The blue line points to the direction of the semi-major axis, while the black arrow indicates the proper motion direction of Crater II derived in this work ($\mu_{\alpha}\cos \delta$=-0.14 $\pm$ 0.07 mas yr$^{-1}$, $\mu_{\delta}$=-0.10 $\pm$ 0.04 mas yr$^{-1}$).}
\label{fig:morph}
\end{figure}

The GB stars (RGB + AGB) of Crater~II overall have an elliptical distribution similar to that found 
for the RRL by \citet{vivas19}, the parameters of which we have adopted for the analysis in this section. 
In Figure \ref{fig:morph} we show the distribution and the morphology for the 863 GB stars from the photometric 
sample together with the proper motion vector obtained in \S~\ref{sec:proper_motion}.

\begin{figure}
\includegraphics[width=0.48\textwidth]{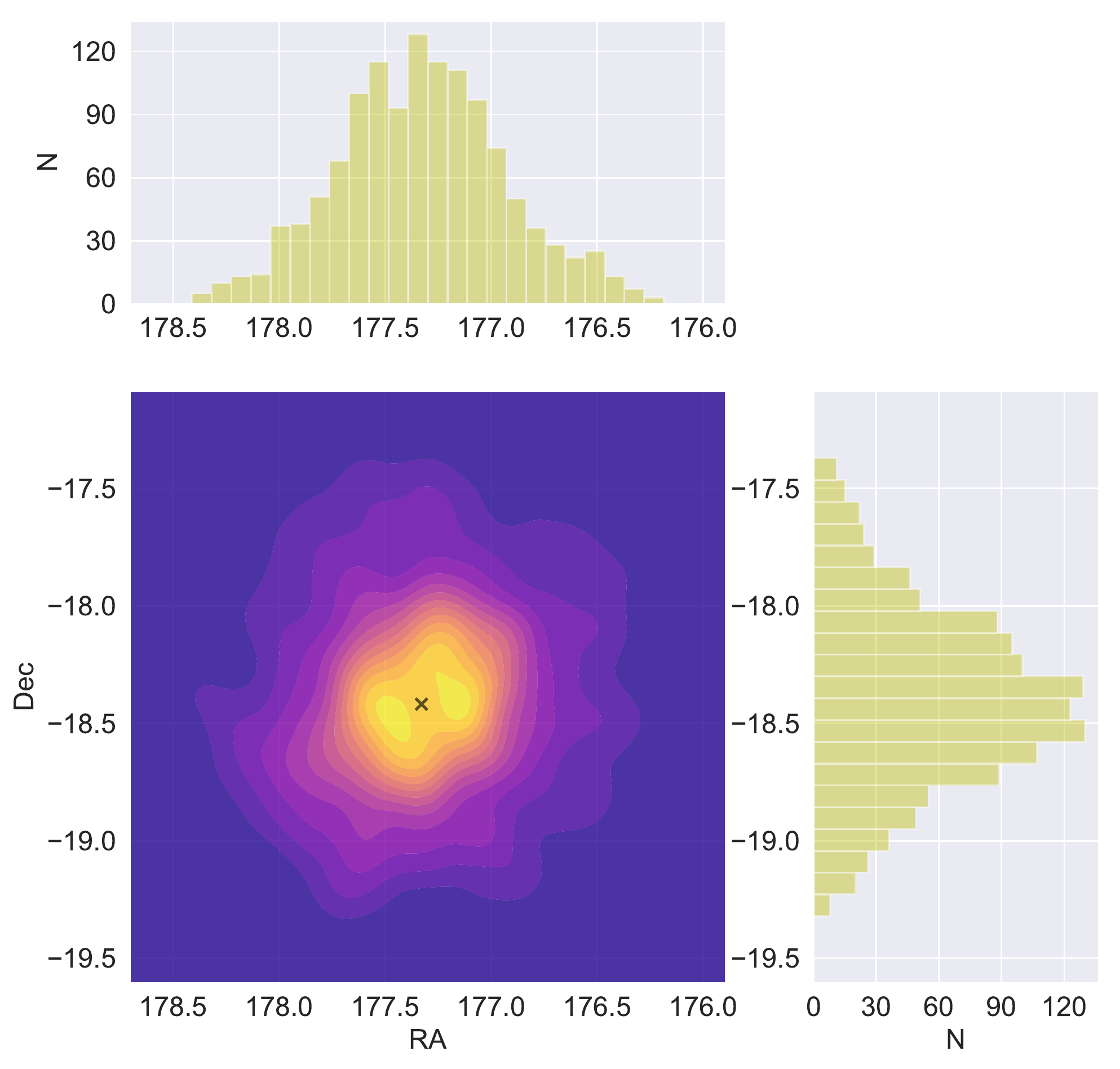}
\caption{Isodensity contour map and RA and DEC distributions of the members (photometric + spectroscopic + variables) of Crater~II. }
\label{fig:isocontourn}
\end{figure}

The distribution of the members of Crater~II (see \S~\ref{sec:cmd_description}) does not show a peaked central concentration, also suggested 
from the distribution of RRL \citep{vivas19} and confirmed by the larger sample of stars here. Although overall the distribution of stars is smooth, on closer examination there are indications of inhomogeneities 
in the spatial distribution of stars. Figure~\ref{fig:isocontourn}, 
an isodensity contour map, shows two overdensities located at (RA = 177.45 deg, DEC = -18.47 deg) and 
(RA = 177.18 deg, DEC = -18.40 deg). They are approximately the same distance from the
Crater~II center (displayed as a black cross), and between these two peaks, in the center of Crater~II, 
there is a 2-$\sigma$ fall in the number of stars. The reliability of these overdensities is statistically confirmed by calculation of Poisson uncertainties.

\begin{figure}
\includegraphics[width=0.48\textwidth]{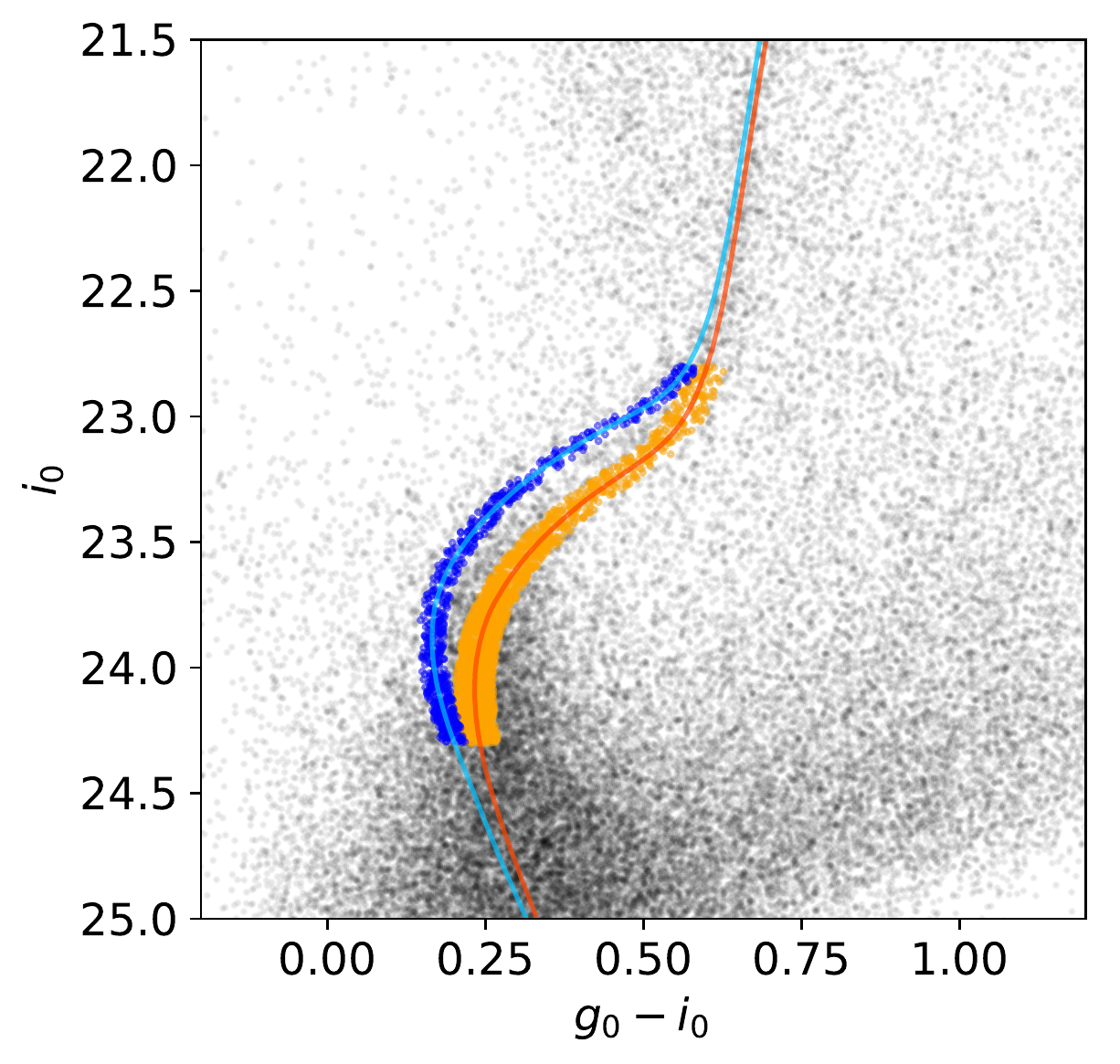}
\caption{Zoom-in of the MS, SGB and lower RGB of the colour-magnitude diagram of Crater~II. Orange (blue) dots represent the older (younger) MS-SGB stars selected under the criteria described in the text.}
\label{fig:cmd_sgb}
\end{figure}

There is an indication from Figure~\ref{fig:cmd} that there may be a difference in the relative numbers of the 
two SGB populations as a function of the elliptical distance (r'), and we investigate this further by selecting stars 
from the SGB and the top of the MS region of the CMD (from $i$ = 22.8 to $i$ = 24.3).
We follow up the 10.5 Gyr isochrone with a circular bin of radius 0.02 mag, and with a circular bin of radius 0.03 mag for the 
12.5 Gyr isochrones.  At the ends of the region under consideration the two tracks are close enough so the stars may appear in both selections, in such cases the closest isochrone is selected so that there is no double counting. The selection performed is shown in Figure~\ref{fig:cmd_sgb}.

We then use the center of Crater~II
and the elliptical shape as defined by the RRL \citep{vivas19} and divide into three annuli, the inner contains 
stars with elliptical distance (r') less than 0.35 degrees, the second with elliptical distances between 0.35 
and 0.65 degrees, and the outer with elliptical distances between 0.70 and 0.78 degrees.  
Although there are Crater~II stars across the whole DECam
field, their numbers in this outer annulus will be very few compared to the field stars, and for the MSTO region 
a good approximation for the purposes of this calculation is zero Crater~II stars, 
however we conservatively 
consider an error of 10 percent in the outer annulus (field) star counts when using this number to correct the inner and center
annulus star counts for field star contamination. For stars definitely identified as 
Crater~II members such as the RRL, no corrections are required for comparing their star counts in the two inner annuli, and the errors quoted derive from the counts only. 

 \begin{table*}
 \begin{center}
 \caption{Radial Distributions of Crater~II Stars}\label{tab:rad_dist}
 \begin{tabular}{cccccccl}
 \hline
 \textit{Population} & \textit{Inner Annulus} & \textit{Center Annulus} & \textit{Outer Annulus}  
 & \textit{Normalised Inner} & \textit{Normalised Center} & \textit{Ratio Inner/Center}\\
  \hline
RR   all    & 43 & 33 &   &     &    & $1.3 \pm0.4$ \\
RRab bright & 12 & 17 &   &     &    & $0.7 \pm0.4$ \\
RRab faint  & 27 & 12 &   &     &    & $2.2 \pm1.1$ \\
12.5 Gyr    & 656 & 692 & 122 & 530 & 383 & $1.38 \pm 0.15$ \\
10.5 Gyr     & 266  & 208  & 32 & 233  & 127 & $1.83 \pm0.15$ \\
Area (deg$^2$) & 0.2925 & 0.7163 & 0.2827 & & & \\

 \hline
 \end{tabular}
 \end{center} 
 \label{tab:raddist}
 \end{table*}

The results in Table~\ref{tab:rad_dist} highlight  the  difference between the central concentration of faint 
and bright RRL, shown as a radial plot in figure 12 of \citet{vivas19}. This has been interpreted as a slightly more metal poor population having a wider spatial distribution than the more metal rich stars. In any case, the spread in metallicity between the two populations is expected to be small. In addition,
we confirm by star counts 
the visual impression that the SGB plus upper MS stars
that follow the 10.5 Gyr isochrone appear slightly more centrally concentrated than those that follow the 
12.5 Gyr 
isochrone.  For more populous dwarf galaxies it is commonly found that younger populations are more centrally
condensed than older populations \citep[see e.g. ][]{harbeck01,tolstoy04,okamoto17}, although generally the age differences are greater than for 
Crater~II.  While it is suggestive here to make an association between the two populations of SG-MSTO
stars and the apparent two groups of RRL stars, a direct association is not supported on stellar evolution 
grounds as discussed above, and the ratios in Table~\ref{tab:rad_dist} have large error bars.  
However, the radial distributions of the RRL and the two SG-MSTO stars are not inconsistent, in a scenario of 
a small increase of metallicity over a rather narrow age range. Spectroscopy of the HB stars, possible with present 
large telescopes, would help to turn the tentative statements from the present data into firmer conclusions.

\section{Crater~II Variable Stars Progenitors}\label{sec:variable_stars}

The RRL variables are discussed by \citet{joo18, monelli18, vivas19}, with the latter paper providing a definitive
analysis of the pulsational properties. The distribution and relation of the RRL to other CMD components are
discussed above.
In addition to the 99 RRL (98 measured from DECam data), 
seven ACs and one DC have been found in Crater~II by \citet{vivas19}.
With the large numbers of blue stragglers (BS) and no young stars, the DC is interpreted as an SX Phoenicis
variable, and its discovery 
amongst the BS stars is not surprising, indeed
there are likely to be more of these (very faint and short period) stars waiting discovery.  For the AC,
there are two production channels \citep{bono97, fiorentino12, cassisi13}, the first is from an intermediate
age population that directly produces the $1-2\ M_{\sun}$ stars, and for which there is no progenitor evidence from
the Crater~II CMD, while the second channel is from BS stars.  This route also has two possible production channels, both
involving merging of two stars; in globular clusters stellar collisions in the dense cluster core will be frequent,
however this route is clearly not significant for Crater~II.  Instead, the coalescence of close 
binaries will be the production channel relevant here \citep{gautschy17}.

\section{Discussion and Conclusions}\label{sec:discussion}

From analysis of a deep CMD, and by comparison with the properties of the RRL discussed in detail in a companion paper \citep{vivas19} we can make a 
number of statements about this unusual, but perhaps not uncommon, galaxy. 

The main new finding in this paper has been possible thanks to the unprecedentedly deep CMD, which reaches the old MSTO with high photometric precision. The stellar populations of Crater~II are characterized by two main events of star formation that occurred at a mean age of 12.5 and 10.5 Gyr ago. 
An old ($\sim 13.5$ Gyr) population seems to be a minor component, as disclosed by the main sequence isochrone fitting, by the HB morphology, which lacks a blue component, and by the characteristics of the RRL variable star population.  These characteristics of the stellar population of Crater~II (M$_V$ = -8.2) are very unusual among UFDs; a similarly extended period of star formation in a UFD has been found so far only in the M31 satellite And~XVI (\citealt{monelli16}; M$_V = -7.3$), even though in that case a double star-formation episode is not evident in the CMD, while it is hinted in the star formation history derived through CMD fitting. What may be the origin of a stellar component with such unusual characteristics? To try to answer this question, we will consider other observed aspects of this galaxy.  

Firstly, the Crater~II proper motion confirmed here, and the orbit that has been derived \citep{fritz18, fu19} suggest strongly that Crater~II should be disrupting due to penetrating well into the more central regions of our Galaxy on each orbit. From the current analysis, we can state, in contrast, that the galaxy structure seems regular, and the younger (10.5 Gyr) population is more centrally concentrated than the older (12.5 Gyr) population, as is common in dwarf galaxies \citep{harbeck01,tolstoy04,okamoto17}, which would support a quiet rather than an episodically violent life. However, we cannot make strong direct statements about this scenario, as a much wider area survey for extra tidal material, in particular RRL, would be needed to provide observational proof. 

In this scenario, a relatively minor old population could be explained by preferential tidal stripping in the earliest pericentric passages (see e.g. \citealt{fritz18} for an orbit integration under different assumptions of the Milky Way potential). The second event of star formation could have occurred or even have been triggered by one of these early pericentric passages, before complete removal of gas would have taken place, while the prior star formation events  are consistent with star formation occurring along the orbit of the galaxy. This is behavior that is expected from simulations (see, e.g. \citealt{nichols15} or \citealt{haus19}) and that has been observed, for example, in the Sagittarius dSph,  which has retained gas and continued star formation during several pericentric passages prior to infalling \citep{siegel07}.

An alternative scenario would be that Crater~II is a relatively recent capture by our Galaxy; a relatively low initial star formation rate could indicate that it formed in an environment isolated from other galaxies.   \citet{gallart15} suggested that the difference between dwarf galaxies with fast evolution (in fast dwarfs, star formation would have started early, probably before reionization and would have terminated early) compared to slow evolution (low intensity of early star formation, which then takes place for most or all of a Hubble time) depended on the density of the environment at time of formation, with fast evolvers forming in high density environments.

 If the low amount of old population in Crater II is intrinsic (that is, not due to preferential stripping of the old population), it implies that the rate of star formation was low prior to reionization,  and by this Crater~II would comply with one of the criteria to be classified as a slow dwarf. However, the relatively early cessation of star formation $\sim$10 Gyr ago is at odds with the normal definition of a slow dwarf. 

A combination of the two scenarios is however plausible. Given the high eccentricity of the derived orbit, and the expected orbital changes along the evolution in live haloes \citep{nichols15}, the early distance of Crater~II to the Milky Way may have been in excess of the currently measured 116.5 kpc \citep{vivas19}, and thus, it would have formed in a relatively isolated environment. In fact, even its currently calculated apocenter is similar to that of Milky Way dSph satellites such as Carina or Fornax that show extended star formation and were classified by slow dwarfs by \citet{gallart15}. Its apocenter is also not unlike the estimated current distance between M31 and And~XVI. Given the more eccentric orbit of Crater~II compared to Carina and Fornax, the closer perigalacticon, possibly combined with the lower mass, would have resulted in a more efficient stripping of the gas in Crater~II than in the other two galaxies, and thus resulted in an earlier cessation of its star formation. Some characteristics betraying the slow nature of Crater~II may still noticeable.  The observational search for stripping remnants mentioned above, and a detailed star formation history, carefully dealing with field star subtraction and modeling the HB and MSTO - SGB regions in particular, will help to better delineate the evolutionary history of this intriguing galaxy.

\section*{Acknowledgements}

We thank Giuseppina Battaglia for helpful discussions. 

This project used data obtained with the Dark Energy Camera (DECam),
which was constructed by the Dark Energy Survey (DES) collaboration.
Funding for the DES Projects has been provided by 
the U.S. Department of Energy, 
the U.S. National Science Foundation, 
the Ministry of Science and Education of Spain, 
the Science and Technology Facilities Council of the United Kingdom, 
the Higher Education Funding Council for England, 
the National Center for Supercomputing Applications at the University of Illinois at Urbana-Champaign, 
the Kavli Institute of Cosmological Physics at the University of Chicago, 
the Center for Cosmology and Astro-Particle Physics at the Ohio State University, 
the Mitchell Institute for Fundamental Physics and Astronomy at Texas A\&M University, 
Financiadora de Estudos e Projetos, Funda{\c c}{\~a}o Carlos Chagas Filho de Amparo {\`a} Pesquisa do Estado do Rio de Janeiro, 
Conselho Nacional de Desenvolvimento Cient{\'i}fico e Tecnol{\'o}gico and the Minist{\'e}rio da Ci{\^e}ncia, Tecnologia e Inovac{\~a}o, 
the Deutsche Forschungsgemeinschaft, 
and the Collaborating Institutions in the Dark Energy Survey. 
The Collaborating Institutions are 
Argonne National Laboratory, 
the University of California at Santa Cruz, 
the University of Cambridge, 
Centro de Investigaciones En{\'e}rgeticas, Medioambientales y Tecnol{\'o}gicas-Madrid, 
the University of Chicago, 
University College London, 
the DES-Brazil Consortium, 
the University of Edinburgh, 
the Eidgen{\"o}ssische Technische Hoch\-schule (ETH) Z{\"u}rich, 
Fermi National Accelerator Laboratory, 
the University of Illinois at Urbana-Champaign, 
the Institut de Ci{\`e}ncies de l'Espai (IEEC/CSIC), 
the Institut de F{\'i}sica d'Altes Energies, 
Lawrence Berkeley National Laboratory, 
the Ludwig-Maximilians Universit{\"a}t M{\"u}nchen and the associated Excellence Cluster Universe, 
the University of Michigan, 
{the} National Optical Astronomy Observatory, 
the University of Nottingham, 
the Ohio State University, 
the OzDES Membership Consortium
the University of Pennsylvania, 
the University of Portsmouth, 
SLAC National Accelerator Laboratory, 
Stanford University, 
the University of Sussex, 
and Texas A\&M University.

Based on observations at Cerro Tololo Inter-American Observatory, National Optical
Astronomy Observatory (NOAO Prop. ID 2017A-0210 P.I. A.R. Walker), which is operated by the Association of
Universities for Research in Astronomy (AURA) under a cooperative agreement with the
National Science Foundation.

This research has been supported by the Spanish Ministry of Economy and 
Competitiveness (MINECO) under the grant AYA2014-56795-P.   CG and MM 
acknowledge support by the Spanish Ministry of Economy and Competitiveness 
(MINECO) under the grant  AYA2017-89076-P.  SC acknowledges support from Premiale INAF MITiC, from INFN (Iniziativa specifica TAsP), and 
grant AYA2013-42781P from the Ministry of Economy and Competitiveness of Spain.

This research has made 
use of the NASA/IPAC ExtraGalactic Database (NED) which is operated by the Jet 
Propulsion Laboratory, California Institute of Technology, under contract with 
the National Aeronautics and Space Administration.


\bsp	
\label{lastpage}
\end{document}